\documentclass[
   ,final            
  ]
  {aipproc}

\layoutstyle{6x9}

\begin{document}

\title{The $g_2$ Structure Function: An Experimental Overview}

                

\author{K. Slifer}{
  address={University of New Hampshire}
}

\begin{abstract}
We will discuss recent results for the spin structure functions, 
with an emphasis on $g_2$.  High precision $g_2$ data allows for tests 
of the Burkhardt--Cottingham sum rule, and is needed to consistently
evaluate higher twist effects. 
\end{abstract}

\maketitle

\section{Introduction}
Four independent structure functions are needed for a complete 
description of nucleon structure.  All spin-dependent effects are contained in $g_1$ and $g_2$, 
while spin-independent effects are parameterized in $F_1$ and $F_2$.
A simple physical interpretation is given in the impulse approximation of the parton model 
as distributions of quark momentum and spin in the nucleon:
\begin{eqnarray}
\nonumber
F_1(x) &=& \frac{1}{2} \sum e_i^2 \left[q_i(x)+ \bar{q}_i(x)\right]\\
F_2(x) &=& 2 x F_1(x) \\
\nonumber
g_1(x) &=& \frac{1}{2}  \sum e_i^2 \Delta q_i(x) 
\end{eqnarray}
Here the summation runs over all quark flavors.

Analysis of the spin structure functions (SSF) is typically facilitated via 
the Cornwall-Norton (CN)
moments~\cite{Cornwall:1968cx,Melnitchouk:2005zr}:
\begin{eqnarray}
\nonumber
\Gamma_1^{(n)}(Q^2) = \int_0^1 dx~ x^{n-1} g_{1}(x,Q^2) \\
\label{eq:CN}
\Gamma_2^{(n)}(Q^2) = \int_0^1 dx~ x^{n-1} g_{2}(x,Q^2)
\label{eq:MOMENTS}
\end{eqnarray}
where $x$ is the Bjorken scaling variable, and $Q^2$ represents the positive definite four momentum transfer of the virtual photon, which mediates the interaction between the incident electron probe and the nucleon target.
By convention, the superscript is usually dropped in the case of $n=1$.

The $Q^2$ and $x-$dependence of the $g_1$ structure function has been measured~\cite{Anthony:2000fn,Anthony:2002hy,Airapetian:2007mh,Amarian:2002ar,Amarian:2004yf,Meziani:2004ne,Osipenko:2004xg,Osipenko:2005nx,Slifer:2008re,Prok:2008ev,Solvigno:2008hk} with impressive precision over a wide kinematic range.  In this proceedings, we focus on the other structure function $g_2$, which has 
been historically neglected due to the
technical difficulties of producing  transversely polarized targets, and
the lack of a simple interpretation of $g_2$ in the classic parton model.

\section{The Burkhardt-Cottingham Sum Rule}


The first CN moment of the $g_2$ structure function is predicted to vanish by the Burkhardt-Cottingham (BC) sum rule~\cite{Burkhardt:1970ti}:
\begin{equation}
\label{eq:bc}
{\Gamma}_2 = \int_0^1 dx~g_2(x,Q^2) = 0
\end{equation}
This sum rule arises from the unsubtracted dispersion relation for the spin-dependent virtual-virtual Compton
scattering  amplitude $S_2$.  It is expected to 
be valid for any value of $Q^2$, although the reader is referred to~\cite{Jaffe:1989xx} for a detailed discussion of scenarios which would lead to violations of the relation.

\begin{figure}
   \includegraphics[height=.5\textheight]{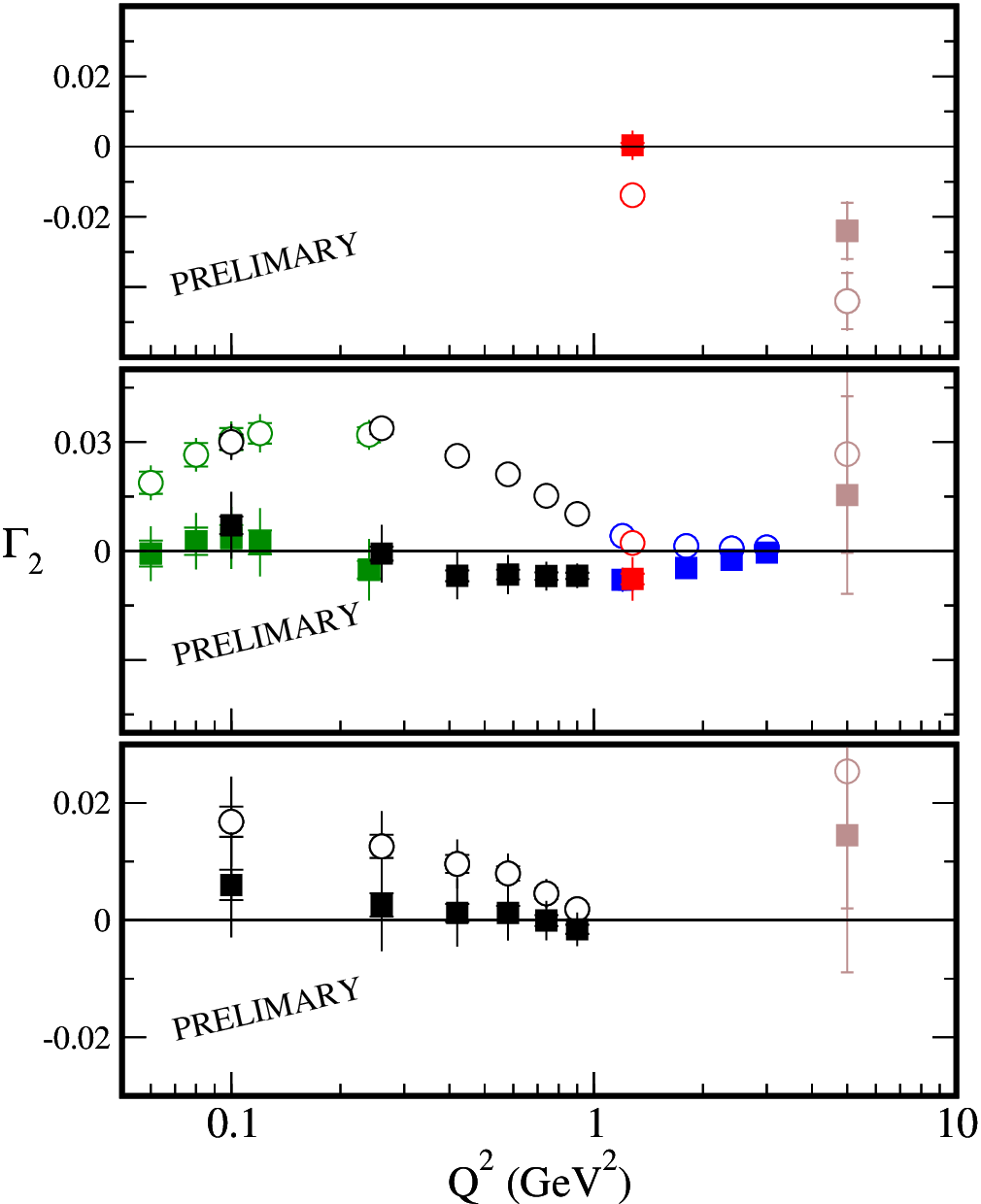}
   \caption{\label{BCWORLD}World data for $\Gamma_2(Q^2)$ for the proton (top), neutron (middle) and $^3$He (bottom). Open symbols represent measured data (typically the resonance region), while the full symbols include an estimate of the unmeasured contributions to the integral. Inner (outer) error bars represent statistical (total) uncertainties. 
Brown: E155 collaboration~\cite{Anthony:2002hy}. 
Red: RSS~\cite{Slifer:2008xu,Wesselmann:2006mw}. 
Black: E94010~\cite{Amarian:2003jy,Slifer:2008re}.
Green: E97110~\cite{SAGDH} (Very Preliminary). 
Blue: E01012~\cite{Solvigno:2008hk} (Very Preliminary).}
 \end{figure}

Fig.~\ref{BCWORLD} displays existing world data for $\Gamma_2(Q^2)$ for the proton, neutron and $^3$He.
All available data are consistent with the nuclear sum rule for $^3$He.  Similarly, we find satisfaction of the neutron sum rule, within uncertainties, across several different experiments and a large range of $Q^2$.  The picture is not so clear for the proton.
The E155 collaboration 
found their data to be inconsistent with the
proton BC sum rule at
$Q^2=5$ GeV$^2$, and the only other data point is from the RSS collaboration at $Q^2 = 1.3$ GeV$^2$.
Two upcoming JLab experiments will greatly expand our knowledge of $g_2^p$
with measurements planned~\cite{DLT,SANE}
in the ranges $0.02<Q^2<0.4$ and $2.5<Q^2<6.5$ GeV$^2$.
%

 \begin{figure} 
   \includegraphics[height=.5\textheight]{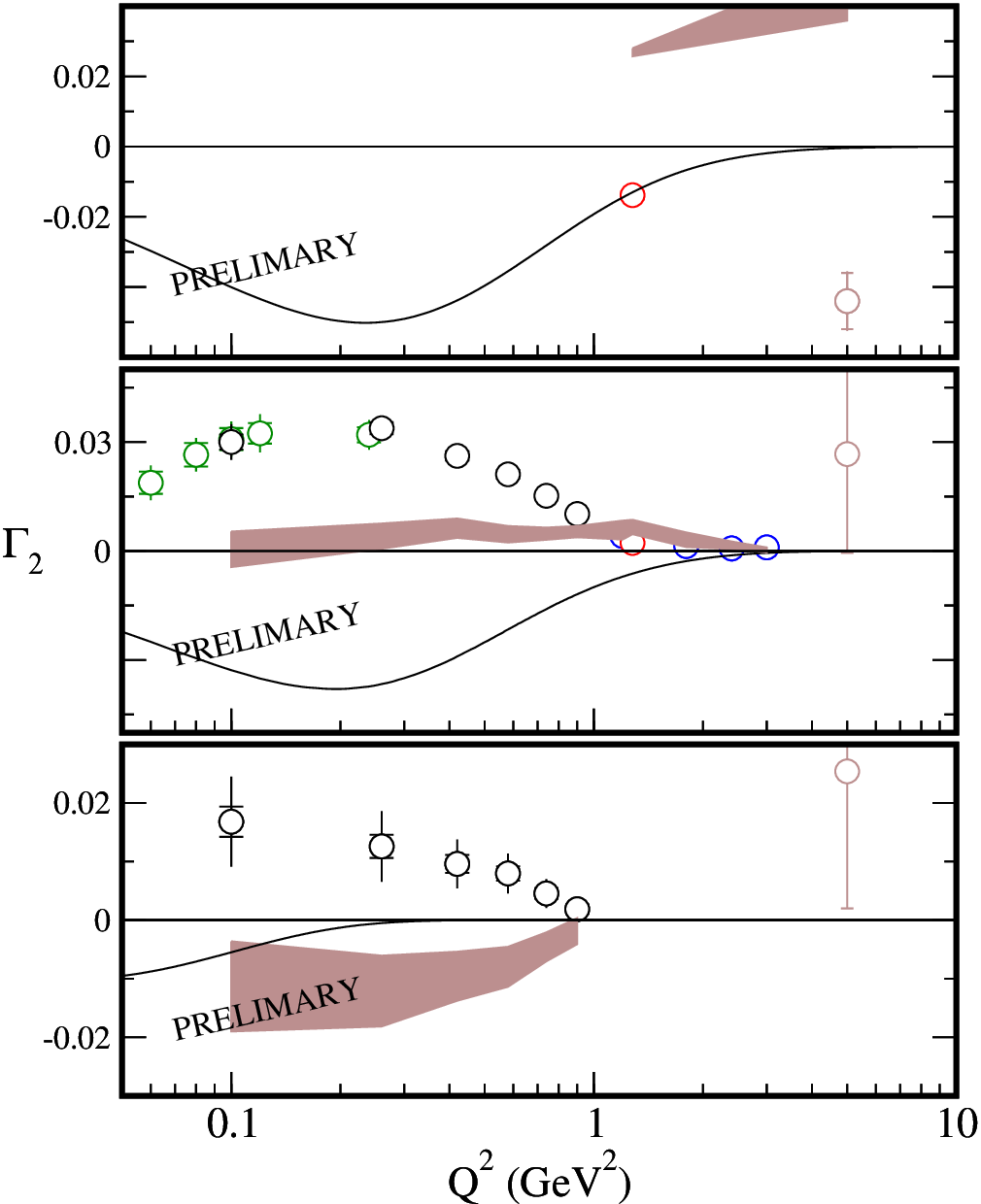} 
  \caption{\label{BCDIS}Full band represents an evaluation of $x\to 0$ contribution to $\Gamma_2(Q^2)$ assuming the validity of the BC sum rule. Proton (top), neutron (middle), and $^3$He (bottom).
Open symbols represent measured data (typically the resonance region). Inner (outer) error bars represent statistical (total) uncertainties.
Brown: E155 collaboration~\cite{Anthony:2002hy}.
Red: RSS~\cite{Slifer:2008xu,Wesselmann:2006mw}.
Black: E94010~\cite{Amarian:2003jy,Slifer:2008re}.
Green: E97110~\cite{SAGDH} (Very Preliminary).
Blue: E01012~\cite{Solvigno:2008hk} (Very Preliminary).
Nucleon form factors from~\cite{Lomon:2002jx,Mergell:1995bf}. Nuclear form factors from~\cite{Amroun:1994qj}.
} 
\end{figure}

The open symbols in Fig.~\ref{BCWORLD} represent the experimentally measured data which typically covers the resonance region.  To evaluate Eq.~\ref{eq:bc}, the contributions from $x=1$ and $x\to 0$ must be included.
The nucleon elastic contribution at $x=1$ can be easily evaluated using the form factor
parametrizations of~\cite{Lomon:2002jx,Mergell:1995bf}, and the nuclear elastic contribution for $^3$He is similarly determined from~\cite{Amroun:1994qj}.  All form factor parameterizations are known to high precision over the relevant region.

To estimate the $x\to 0$  contribution for the $g_2$ integrals,  
the Wandzura-Wilczek~\cite{Wandzura:1977qf} relation is used:
\begin{eqnarray}
\label{eq:WW}
g_2^{WW}(x,Q^2) \equiv -g_1(x,Q^2) + \int_x^1 \frac{g_1(y,Q^2)}{y} dy
\end{eqnarray}
This relation gives a prediction for the leading twist behaviour of $g_2$ entirely in terms of $g_1$.  Using Eq.~\ref{eq:WW} for the low $x$ contribution of the integral relies on the assumption that higher twist effects asymptotically vanish as $x\to 0$, which is, of course, untested due to the difficulty in measuring structure functions in this region.   This contribution to the integral represents the largest contribution to the overall systematic uncertainty.

\subsection{What can the BC sum rule tell us about low $x$?}
Due to the extreme low counting rate, precise $g_2$ data at low $Q^2$ and low $x$ are unlikely to be obtained anytime in the near future.  However, if one were to accept the BC sum rule purely on theoretical grounds, it is possible to invert Eq.~\ref{eq:bc} to provide information on $g_2$ in this region.  The high precision of the elastic form factors and the measured resonance region data allows us to determine the $x \to 0$ contribution to $\Gamma_2(Q^2)$ with relatively small uncertainty.  Fig.~\ref{BCDIS} shows the resonance region (open symbols) and elastic contribution (solid line) to Eq.~\ref{eq:bc}.  The solid band is the  $x \to 0$ contribution inferred by assuming that Eq.~\ref{eq:bc} vanishes.

\subsection{Violations of the BC Sum Rule?}
As a necessary caveat to the preceding discussion, we stress that the experimental verification of the BC sum rule is still in question. In particular, Table~\ref{tab:sd} gives some perspective on the relative uncertainty of the determinations of $\Gamma_2^n$ in the vicinity of $Q^2\approx 1.0$ 
GeV$^2$.  Satisfaction of the neutron BC sum rule is found only at the 2 to 2.5 $\sigma$ level of the 
total uncertainty.  As can be seen from the table, the uncertainty is typically dominated by systematic effects, while the statistical uncertainties are quite small.  The dominant systematics of E94010 arose from the radiative corrections.  
Considering the significant improvements in knowledge of structure functions at low $Q^2$ 
that has occurred in recent years, a re-examination of this uncertainty is warranted in order to provide a conclusive test of the BC sum rule.

 \begin{figure}
   \includegraphics[height=.5\textheight]{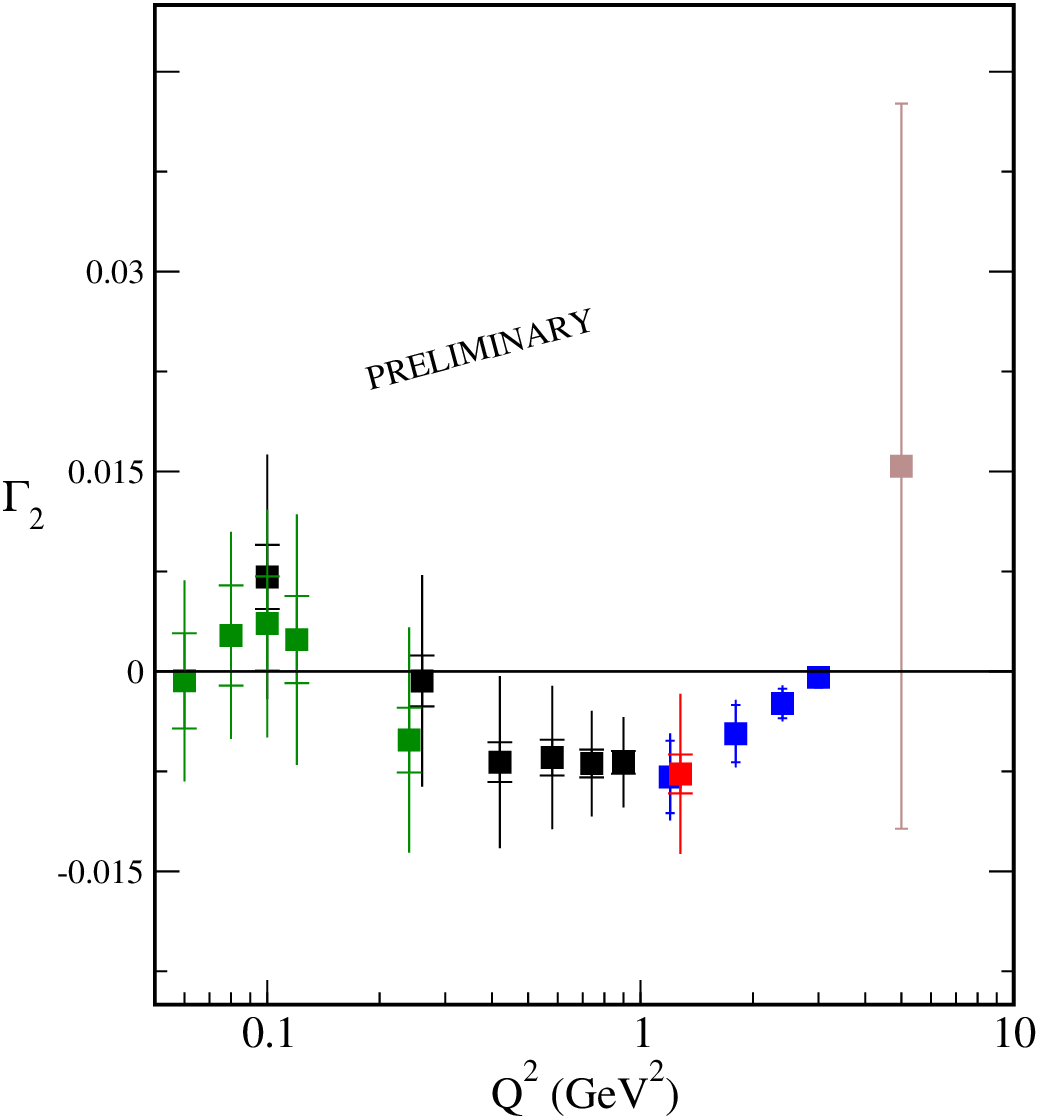}
  \caption{\label{BCZOOM}$\Gamma_2^n(Q^2)$. Full integral for neutron.
Inner (outer) error bars represent statistical (total) uncertainties.
Brown: E155 collaboration~\cite{Anthony:2002hy}.
Red: RSS~\cite{Slifer:2008xu,Wesselmann:2006mw}.
Black: E94010~\cite{Amarian:2003jy,Slifer:2008re}.
Green: E97110~\cite{SAGDH} (Very Preliminary).
Blue: E01012~\cite{Solvigno:2008hk} (Very Preliminary).
}
\end{figure}

\begin{table}
\begin{tabular}{ccc}
\hline
$Q^2$(GeV$^2$) &Total Relative Uncertainty & Statistical Relative Uncertainty\\ 
\hline
0.74 & 1.7 & 6.6 \\
0.90 & 2.0 & 7.9 \\
1.20 & 2.4 & 2.9 \\
1.80 & 1.9 & 2.2 \\
2.40 &  1.8 & 2.2 \\
\hline
\end{tabular}
\caption{Standard deviations of $\Gamma_2^n(Q^2)$ from zero. From E94010~\cite{Amarian:2003jy} for $Q^2<1.0$ GeV$^2$ 
and E01012~\cite{Solvigno:2008hk} (Very Preliminary) for  $Q^2>1.0$ GeV$^2$} 
\label{tab:sd}
\end{table}

\section{Higher Twist Measurements}

The Operator Product Expansion
(OPE)~\cite{Wilson:1969zs,Kodaira:1994ge,Ehrnsperger:1993hh}
relates the structure functions to QCD quark and gluon matrix elements,
which contain information on parton interactions.
This effective theory of  QCD is formulated in terms of the CN moments of Eq.~\ref{eq:CN}.
The OPE provides the following expansion of  the lowest moment $\Gamma_1(Q^2)$:
\begin{eqnarray}
\Gamma_{1}(Q^2) = \sum_{\tau=2,4,\ldots} \frac{\mu_\tau(Q^2)}{Q^{\tau-2}}
\label{eq:MOMENTS2}
\end{eqnarray}
This expansion is performed in inverse powers of the \emph{twist} $\tau$.
Successful predictions from the parton model map onto the leading twist of the OPE, while deviations from leading twist behaviour are known as higher twist effects.
The parton model says nothing about the $g_2$ structure function, making it in an excellent quantity 
to study higher twist.

The coefficients $\mu_\tau$ are identified with nucleon matrix elements of QCD operators of twist $\le\tau$.  The first coefficient beyond leading twist is $\mu_4$, which can be decomposed as:
\begin{eqnarray}
\mu_4 = \frac{1}{9}M^2\left(\tilde{a}_2+4\tilde{d}_2 + 4\tilde{f}_2\right)
\end{eqnarray}
This involves a combination of the leading twist, twist-3 and twist-4 matrix elements, respectively.

\subsection{Higher Moments}
\begin{figure}
  \includegraphics[height=.5\textheight]{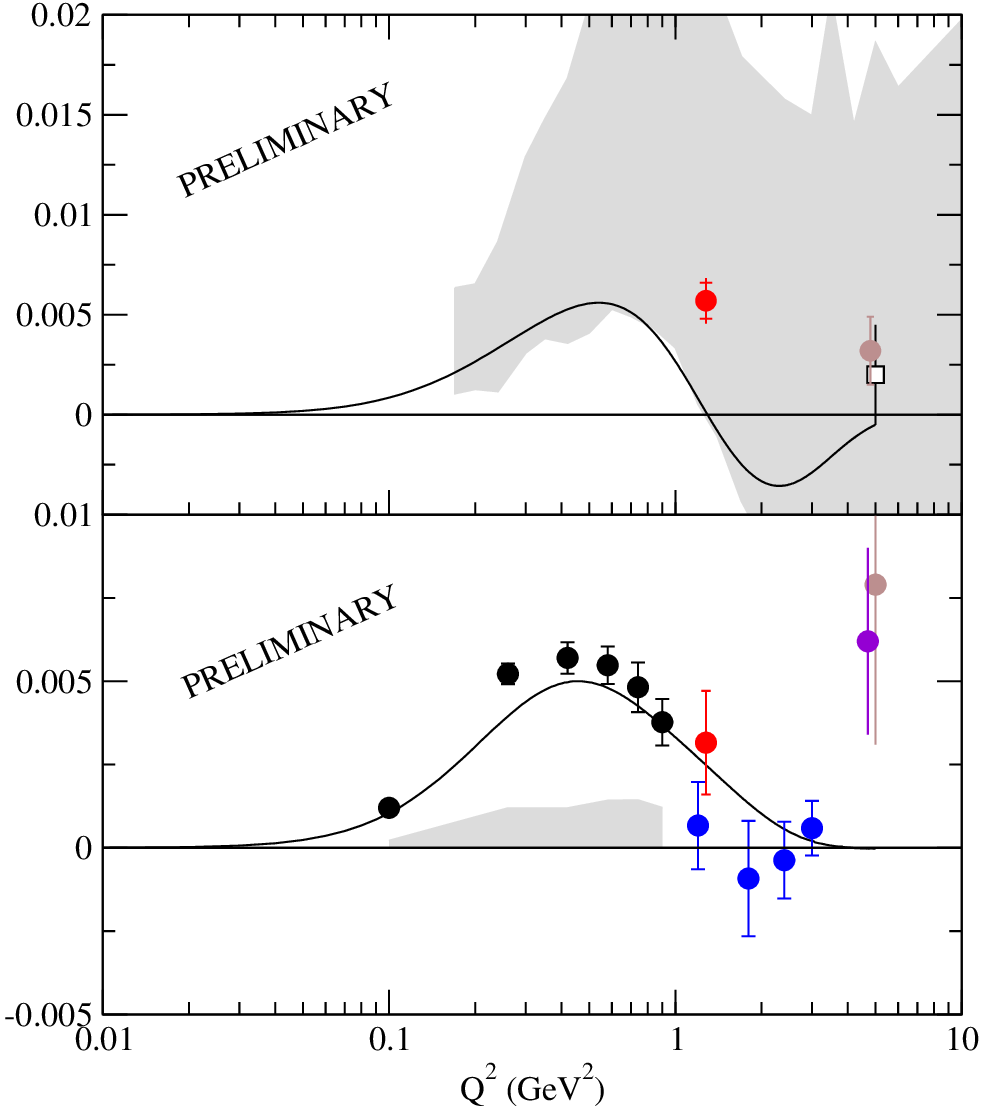}
  \caption{\label{fig:d2}$I(Q^2)$ for proton (top)  and neutron (bottom).
Brown: E155 collaboration~\cite{Anthony:2002hy}.
Black open square: QCDSF~\cite{Gockeler:2000ja}
Red: RSS~\cite{Slifer:2008xu,Wesselmann:2006mw}.
Black circle: E94010~\cite{Amarian:2003jy}.
Green: E97110~\cite{SAGDH} (Very Preliminary).
Blue: E01012~\cite{Solvigno:2008hk} (Very Preliminary).
Magenta: E99117~\cite{Zheng:2004ce}.
Solid line: MAID model~\cite{Drechsel:1998hk}.
In the top panel, the filled band represents a determination of $I(Q^2$) based on existing world data~\cite{Osipenko:2005nx}.  The large uncertainty reflects the lack of knowledge of $g_2^p$ in this region.
}
\end{figure}

The twist-2 term $\tilde{a}_2$ is purely kinematical in nature and can be evaluated~\cite{Dong:2008zz} from the (target mass corrected) third moment of $g_1$:
\begin{eqnarray}
\label{eq:a2}
\int_0^1 x^2 g_1(x,Q^2) dx = \frac{1}{2}\tilde{a}_2 + \mathcal{O}\left(\frac{M^2}{Q^2}\right)
\end{eqnarray}
Similarly, the third moment of  $g_1$ provides information on $\tilde{d}_2$:
\begin{eqnarray}
\label{eq:d2a2}
\int_0^1 x^2 g_2(x,Q^2) dx = \frac{1}{3}(\tilde{d}_2 - \tilde{a}_2) + \mathcal{O}\left(\frac{M^2}{Q^2}\right)
\end{eqnarray}

Eqs.~\ref{eq:a2} and~\ref{eq:d2a2} are often combined to extract the twist-3 matrix element via:
\begin{eqnarray}
I(Q^2) &=& \int_0^1 dx~ x^2\left(2 g_1 + 3 g_2\right) \\
      &=& \tilde{d}_2(Q^2) + \mathcal{O}\left(\frac{M^2}{Q^2}\right)
\end{eqnarray}
However, note that $I(Q^2)$ is not equivalent to the twist-3 matrix element $\tilde{d}_2$ in any kinematic region where $M^2$ is not negligible compared to $Q^2$~\cite{Dong:2008zz}.  World data on $I(Q^2)$ is shown in Fig.~\ref{fig:d2}. In the top panel, the filled band represents a determination of $I(Q^2$) based on existing world data~\cite{Osipenko:2005nx}.  The large uncertainty reflects the lack of knowledge of $g_2^p$ in this region.

\section{Target Mass Corrections}

\begin{figure}
  \includegraphics[height=.5\textheight]{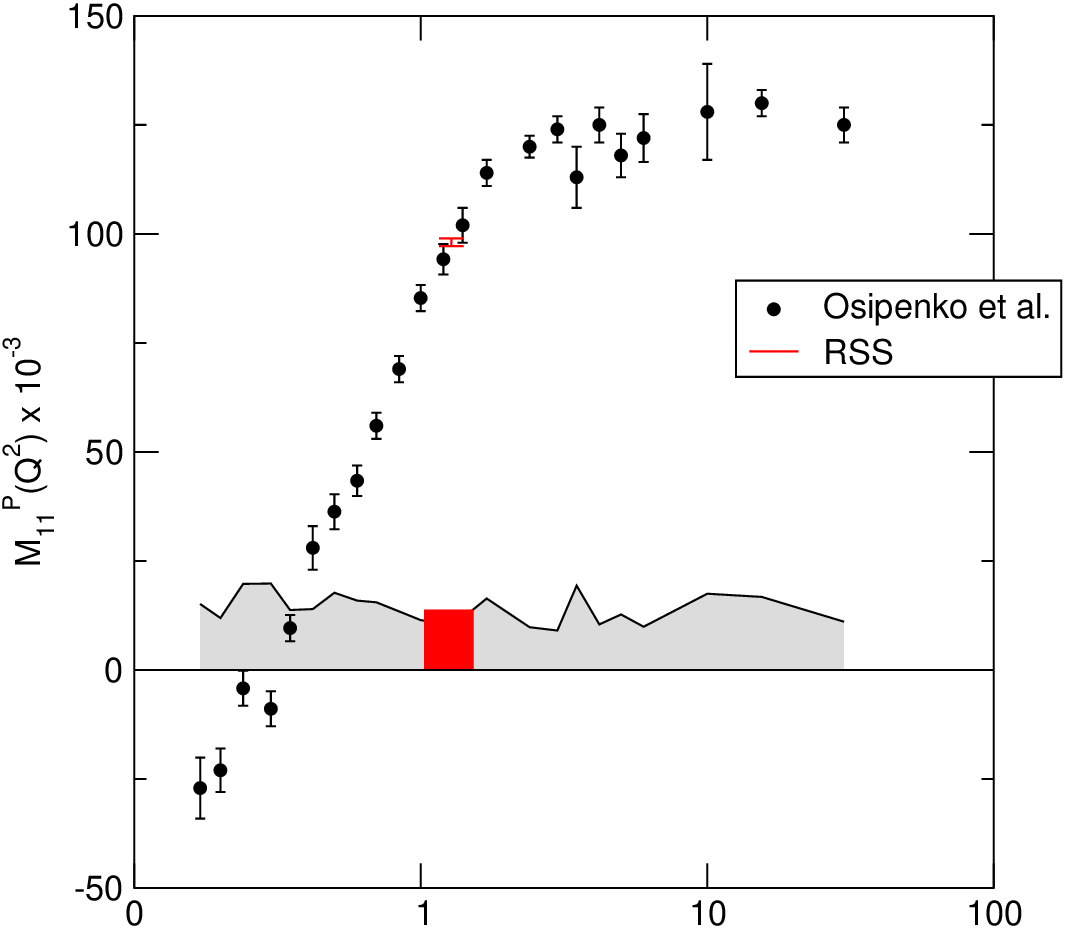}
  \caption{\label{fig:M11}$M_{1}^1(Q^2)$ from~\cite{Osipenko:2005nx} and  RSS~\cite{Slifer:2008xu,Wesselmann:2006mw} (Preliminary). Statistical errors are shown on the data points, while systematic errors are represented by the full bands on the horizontal axis.}
\end{figure}

\begin{figure}
  \includegraphics[height=.5\textheight]{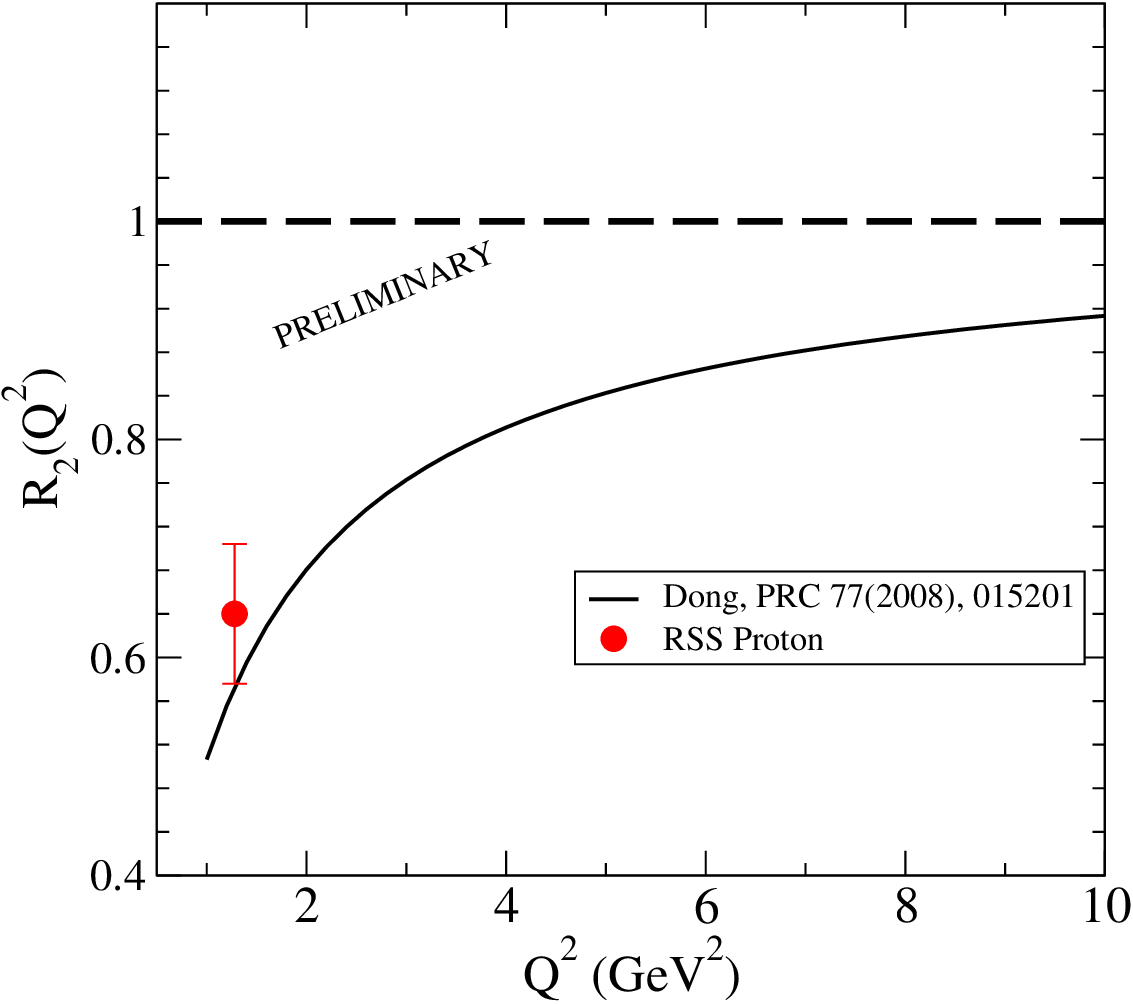}
  \caption{\label{fig:R2} RSS (very preliminary)  $R_2$ data compared to the prediction of Ref.~\cite{Dong:2008zz} }
\end{figure}

As emphasized in~\cite{Dong:2008zz}, dynamical higher twists can be extracted
from the measured SSFs
by using the Nachtmann
moments~\cite{Matsuda:1979ad,Piccione:1997zh}, which
utilize the variable
$\xi= 2x/(1+\sqrt{1+(2xM)^2/Q^2})$. 
The moments are defined:
\begin{eqnarray}
\label{eq:NACMOMENTS}
\nonumber
M_1^n(Q^2)&\equiv& \frac{\tilde{a}_{n-1}}{2} \equiv \frac{1}{2}a_nE_1^n(Q^2,g)=
\int_0^1 \frac{dx}{x^2}\xi^{n+1}\\
\Bigl[
\Bigl\{\frac{x}{\xi}&-&\frac{n^2}{(n+2)^2}\frac{M^2 x\xi}{Q^2}  \Bigr\} g_1
-
\frac{4n}{n+2}\frac{M^2 x^2}{Q^2}g_2
\Bigr]~~~~~\\
\label{eq:NACMOMENTS2}
\nonumber
M_2^n(Q^2)&\equiv&\frac{\tilde{d}_{n-1}}{2} \equiv \frac{1}{2}d_nE_2^n(Q^2,g)=
\int_0^1 \frac{dx}{x^2}\xi^{n+1}\\
&&\Bigl[
\frac{x}{\xi} g_1
+
  \Bigl\{\frac{n}{n-1}\frac{x^2}{\xi^2}-
  \frac{n}{n+1}\frac{M^2x^2}{Q^2}\Bigr\}g_2
\Bigr]~~~~~
\end{eqnarray}

In the limit $M^2/Q^2\to 0$, (i.e. TMC vanish)
the Nachtmann moments reduce to the
more familiar CN moments:
\begin{eqnarray}
M_1^1(Q^2) &\to&  \Gamma_1(Q^2) \\
2 M_2^3(Q^2) &\to& I(Q^2)
\end{eqnarray}
At finite $Q^2$, the ratio of Nachtmann to CN moments gives a quantitative
measure of the TMC:
\begin{eqnarray}
R_1(Q^2)&=&\frac{M_1^1(Q^2)}{\Gamma_1(Q^2)} \\
R_2(Q^2) &=& \frac{2M_2^3(Q^2)}{I(Q^2)}
\end{eqnarray}

The Nachtmann moments allow a clean separation of dynamical higher twists.
In particular, Ref.~\cite{Dong:2008zz} shows that $M_2^3$ gives access to the twist-3 matrix element that is accurate to $\mathcal{O}(M^8/Q^8)$.

A comparison of RSS data for $M_1^1$ to the analysis of Ref.~\cite{Osipenko:2005nx} is shown in
Fig.~\ref{fig:M11}.
Ref.~\cite{Dong:2008zz} predicts that experimental determinations of $I(Q^2)$ significantly overestimate contributions to the twist-3 matrix element for $Q^2< 10$ GeV$^2$.  Fig.~\ref{fig:R2} provides support for this conclusion from preliminary RSS proton data.

\section{Conclusion}
We have presented an overview of existing tests of the Burkhardt--Cottingham sum rule.  
All  neutron and $^3$He data support the sum rule prediction within experimental uncertainties, although the agreement is only at the 2.5$\sigma$ level for the neutron. Proton data is found to be lacking.  We have also presented a Nachtmann moment analysis of existing proton $g_2$ data, which shows that CN moments can significantly overestimate determinations of the twist-3 matrix element.


\begin{theacknowledgments}
I would like to thank the spokesmen of RSS: Mark Jones and Oscar Rondon,  E97110, 
Alexandre Deur, J.P. Chen and Franco Garibaldi,  and E01012: Nilanga Liyanage, J.P Chen and Seonho Choi for kindly allowing me to present their preliminary data here.  Many thanks also to Vince Sulkosky and Patricia Solvignon for providing the raw data.
I would also like to thank Oscar Rondon for many instructive discussions on the topic material.
\end{theacknowledgments}

\bibliographystyle{aipproc}   

\bibliography{karl_slifer}

\end{document}